Towards Global Earthquake Early Warning with the MyShake Smartphone Seismic Network

Part 1 - Detection algorithm and simulation platform

Authors: Qingkai Kong, Robert Martin-Short, Richard M. Allen

Corresponding author: Qingkai Kong

Address: 209 McCone Hall, UC Berkeley, Berkeley, CA, 94720

Email: kongqk@berkeley.edu

**Abstract:**

The MyShake project aims to build a global smartphone seismic network to facilitate large-scale earthquake early warning and other applications by leveraging the power of crowdsourcing. The MyShake mobile application first detects earthquake shaking on a single phone. The earthquake is then confirmed on the MyShake servers using a "network detection" algorithm that is activated by multiple single-phone detections. In this part one of the two paper series, we present a network detection algorithm and a simulation platform to test earthquake scenarios at various locations around the world. The proposed network detection algorithm is built on the DBSCAN classic spatial clustering algorithm, with modifications to take temporal characteristics into account and the association of new triggers. We test our network detection algorithm using real data recorded by MyShake users during the M4.4 January 4th, 2018, Berkeley and the M5.2 June 10th, 2016, Borrego Springs earthquakes to demonstrate the system's utility. In order to test the entire detection procedure and to understand the first order performance of MyShake in various locations around the world representing different population and tectonic characteristics , we then present a software platform which can simulate earthquake triggers in hypothetical MyShake networks. Part two of this paper series explores our MyShake early warning simulation performance in selected regions around the world.

**Introduction**

Earthquake Early Warning (EEW) is a technology that uses networks of seismometers to quickly determine the location and magnitude of an earthquake after it has begun and issues warnings to regions anticipated to experience shaking (e.g. Kanamori, 2007; Allen *et al.*, 2009; Allen and Melgar 2019). Such alerts are typically sent within seconds of the earthquake origin time and can provide up to several minutes of warning depending on the geometry of the monitoring network and the distance between the event and population centers (Allen, 2011, 2013). During this warning time, actions can be taken by individuals and organizations that could potentially save lives and mitigate damage (Strauss and Allen, 2016). In order to be effective, EEW requires the existence of a dense seismic network that has the capability of real-time monitoring of potential earthquake signals. The closer the instruments are to the epicenter, the faster the detection and hence the larger the warning times can be. EEW has been mainly developed using traditional seismic and geodetic networks, which are costly to operate and only exist within a small number of countries (Allen and Melgar 2019). Much of the global population at high risk from earthquake damage thus currently is not benefitting from EEW.

Many alternative, cheaper, non-traditional networks have been proposed, including microelectromechanical system (MEMS) accelerometers installed in buildings, USB accelerometers attached to personal computers or other low-cost sensory equipment (Cochran *et al.*, 2009; Luetgert *et al.*, 2009; Chung *et al.*, 2011; Clayton *et al.*, 2015; Wu, 2015; Wu *et al.*, 2016; Nugent, 2018). Although promising, these ideas suffer from the same disadvantages as traditional networks in that they require physical installation and maintenance by the network operators, which hampers the sustainability and expandability of the EEW system, especially in remote regions.

Recent advances in mobile accelerometer technology mean that smartphones are becoming a viable alternative to fixed seismometers as the primary sensing instruments for EEW (Faulkner *et al.*, 2011; Dashti *et al.*, 2012; Finazzi, 2016; Kong, Allen, Schreier, *et al.*, 2016). Furthermore, there is also interest in the development of the smartphone networks that use GPS and users' mobile application launching times to detect earthquakes (Minson *et al.*, 2015; Bossu *et al.*, 2018; Steed *et al.*, 2019). There are many advantages of using smartphone networks for this application: The devices are globally ubiquitous, even in regions without traditional earthquake monitoring. Since the hardware is maintained by the users, the only requirement for the network operators is to develop and market a software application that can be made accessible via the Google Play or iOS store, and then to maintain a cloud server to collect data. This makes the network easier to maintain and grow.

However, the use of smartphones for earthquake early warning is not without its challenges. Namely, the detection software must be capable of reliably distinguishing between earthquake shaking and all other vibrations that the device might experience. Furthermore, the noise floor of mobile accelerometers is significantly higher than that of traditional seismometers, the extent of coupling between the smartphone and the ground may be poor and the recording of earthquakes is not a priority for users.

MyShake is a smartphone application developed at the UC Berkeley Seismology Lab to monitor smartphone accelerometer data and detect earthquakes. It uses an artificial neural network (ANN) trained on examples of earthquake and non-earthquake waveforms and is able to successfully distinguish earthquake motions from human activity-related motion recorded by the phone (Kong, Allen, Schreier, *et al.*, 2016; Kong, Inbal, *et al.*, 2019). The MyShake application monitors the accelerometer on the device and sends real-time messages containing time, location and ground acceleration data to a server when earthquake-like motions are detected. Kong, Allen, Schreier, *et al.*, (2016) and Kong, Inbal, *et al.*, (2019) should be consulted for a complete description of the MyShake application and its operation. Since the app's first public release in February 2016, MyShake phones have successfully recorded over 900 earthquakes worldwide, the app has approximately 300,000 downloads and 40,000 active users, with approximately 6000 devices making data contributions daily. The data recorded by MyShake has potential uses for various applications, such as mapping ground motion (Kong, Allen, and Schreier, 2016), routine seismic operation (Kong, Patel, *et al.*, 2019), building health monitoring (Kong *et al.*, 2018), and dense array detection (Inbal *et al.*, 2019).

EEW is also a goal of this global smartphone seismic network. However, due to the fact that the current network is relatively sparse, especially outside the US, the potential for MyShake networks to contribute to EEW has not been systematically assessed beyond a handful of basic simulations. Such systematic assessment is vital before MyShake can begin to issue public early warnings. The usefulness of MyShake networks for early warning will vary from region to region, depending on a wide range of factors such as the distance between population centers and active faults, the density and distribution of MyShake users and the origin time, and the magnitude of the earthquake. Quantification of these factors will allow the MyShake development team to identify regions of the world where earthquake early warning with MyShake would be feasible and most beneficial, the minimum number or density of users required for accurate rapid detections, and the likely warning times that could be issued to major population centers in the event of large earthquakes.

This part one of the two paper series describes a simulation platform that has been built to understand the performance of MyShake networks. The platform, built on top of MyShake observations with the aid of a simple physics model and a series of machine learning algorithms, can be used to test and understand the whole MyShake workflow from individual phone triggers to the final detection of the earthquake and estimation of the alerting area. It can simulate the trigger times and ground acceleration values that might be expected from hypothetical MyShake networks responding to given input events and population densities. The locations, times and ground motions reported by individual phones are provided to a network detection algorithm, which first determines whether or not an earthquake is occurring and then uses the trigger information to estimate the earthquake's location and magnitude. Once the earthquake has been located, the system estimates the radius of the region expected to experience shaking of intensity >=MMI 4, for which a warning could be issued. As the simulation proceeds, the earthquake hypocenter parameters are updated as more trigger information becomes available.

We test our network detection workflow on real data collected from devices running MyShake during the June 2016 M5.2 event in Borrego Springs, CA and January 2018 M4.4 event in Berkeley, CA, which are currently the locations with the highest density of MyShake users. Had the system been operating at the time, it could have provided about 6 seconds of warning before the arrival of strong shaking at Palm Springs, and several seconds to much of the San Francisco bay area. This confirms the ability of MyShake networks to issue useful early warnings.

Following this test with real data, we conduct simulations for all historical earthquakes M > 4.0 since January 1$^{st}$ 1980 for a range of earthquake-prone regions around the world including California, New Zealand, Nepal, Central America, Haiti and Sulawesi (Indonesia) in addition to several others shown in the supplementary material. These simulations are described in part two of this two-paper series .

**Overview of the simulation platform**

Our simulation platform consists of several components, shown in Figure 1. These components can be divided into two main functionalities. The first is a mechanism for simulating MyShake-phone networks and their response to earthquakes. It is supplied with the coordinates of the region of interest, the proportion of the population of that region assumed to have the MyShake app installed and the parameters of the earthquakes to be simulated (i.e. location, origin time, magnitude). At each timestep of the simulation, ground acceleration values at each device are estimated and used to determine if the device will 'trigger'. Although hypothetical by nature, this simulation function builds upon observations of real MyShake networks in order to set thresholds for device triggering in addition to uncertainties in the reported times and acceleration values.

The second functionality is a network detection algorithm that takes trigger times, locations and ground acceleration values from the triggered phones and uses them to 1) determine if the network is experiencing an earthquake and 2) if an earthquake is occurring, estimate its location, origin time and magnitude as quickly as possible. This algorithm can be run on data received by real MyShake devices as in our Borrego Spring and Berkeley test cases, or from simulated triggers.

In this section we describe each component of the MyShake simulation workflow, beginning with how we sample the population, through simulating triggers on a device level to the details of our network detection algorithm. For each component we note the parameters that can have a significant impact on the results and justify our choice of their default values.

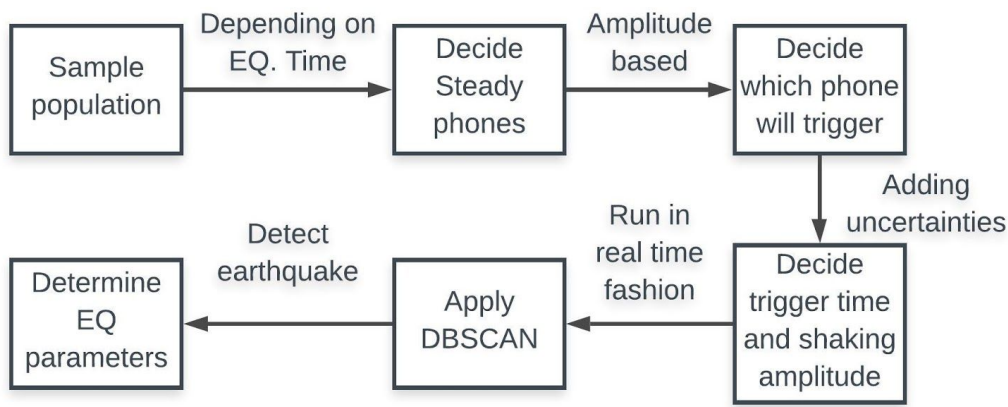

**Figure 1.** Workflow for the MyShake simulation platform

*Sampling the population*

Prior to each simulation, we need to determine the spatial distribution of the simulated MyShake network. This is done by inputting the fraction of the population of the region of interest assumed to have the MyShake application installed on their mobile device. Currently we use 0.1% as a default. User locations are then found by randomly sampling cells of a 1km by 1km grid within the area of interest with a sampling probability weighted by the population in that cell. Once a cell has been identified, the coordinates of the simulated device are drawn from a uniform distribution within the cell. The world population data is obtained from the 2015 Gridded Population of the World, Version 4 (GPWv4). This procedure allows for random sampling of the population while also taking density into account, naturally leading to a greater density of simulated devices in urban areas.

*Identifying stationary phones*

In the current MyShake deployment, a phone must be stationary for 30 minutes before it starts to monitor for earthquake shaking. We use a relationship based on data from the existing global MyShake network to estimate the proportion of active devices that are 'steady' given the origin time of the event to be simulated. The proportion of MyShake devices that are steady for more than 30 min varies significantly over each 24 hour cycle, reflecting the temporal, dynamic nature of the network and is shown in Figure 2. Figure 2a indicates that the network has more phones steady for detecting earthquakes at night than in the daytime. This difference is encoded into the simulation platform.

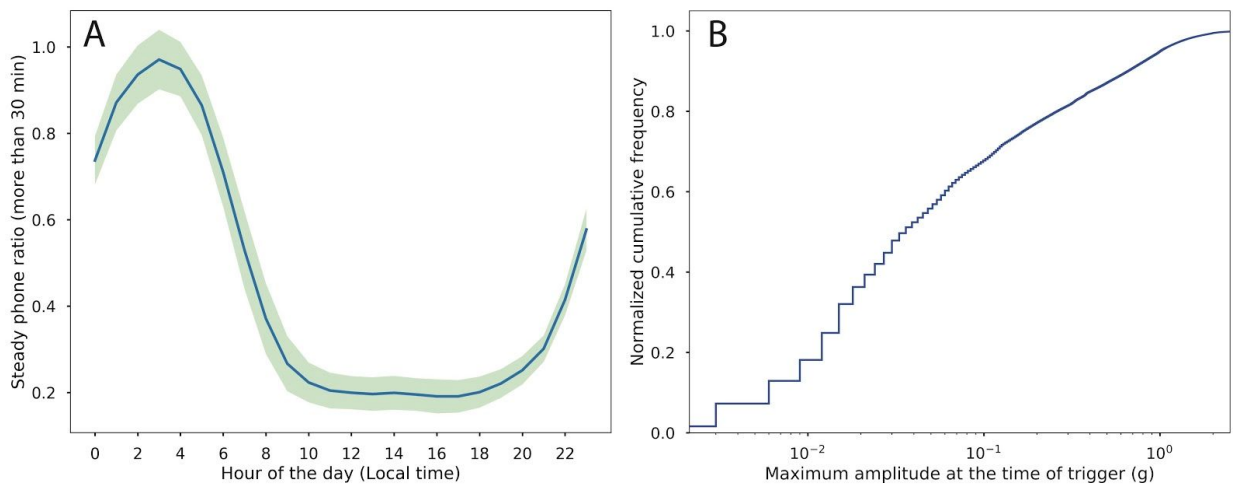

**Figure 2.** Panel A shows the percentage of phones that are steady (stationary) for more than 30 minutes during each hour of the day. The solid line is the average percentage, while the shaded area is the standard deviation. The data shown here was obtained from MyShake users between 2017-07-01 to 2018-07-01 (modified from figure 7b in Kong, Inbal, *et al.*, 2019). Panel B shows the normalized cumulative frequency of the amplitude value at the time of the trigger from 10,377,964 MyShake recordings.

*Determining which phones will trigger*

The triggering mechanism of individual phones for the current simulation platform is determined by an amplitude based approach. Ground motion accelerations associated with P and S-waves at each device are estimated using the distance-magnitude relationships developed by Cua and Heaton (2009), which, given an event magnitude and distance, return the mean and standard deviation of the estimated ground motion distribution for P- and S-waves separately. These relationships are empirical, based on observations from earthquakes in southern California and known to saturate for events of M>6.5 (Cua and Heaton, 2009). Nevertheless, they provide a convenient and relatively accurate way for us to estimate ground motion values.

The ground motion associated with each device is then sampled from the normal distribution returned by the Cua & Heaton (2009) relationships. If the reported ground acceleration value at a device exceeds 0.01g, the phone is assigned a triggering probability of 0.8. Below this threshold, the phone will have a triggering probability defined by p = amplitude / 0.01. This 0.01g threshold was determined via the observation that more than 80% of phone triggers from the real MyShake network have amplitudes larger than 0.01g at the trigger time (Figure 2b).

Analysis of real MyShake triggers has indicated that it is possible to discriminate between whether the phone has triggered on a P- or S-phase by calculating the ratio between the maximum amplitude recorded on the vertical component and the maximum value on the horizontal components in a 2-second window around the trigger. Our tests indicate that this phase discrimination procedure has an accuracy of 70%, so in the trigger simulation workflow it labels the phase of the pick with a 70% accuracy. Random triggers are also simulated at a rate determined by the triggering rate from the data collected by the MyShake network at each hour of the day. Then at each timestamp we use the random triggering rate to determine the number of phones in the region that will also send triggers that are not caused by the earthquake and have random amplitudes.

*Determining phone trigger time and shaking amplitude*

We assume constant P- and S-wave velocities of 6.10 and 3.55 km/s respectively in a half-space model, which allows us to determine travel times of phases to each device. This is clearly a simplification of the true velocity structure but it gives us a first order estimation of the performance of the system. In order to account for uncertainties in the observed trigger times from real events due to poor clock accuracy, or the phone not triggering on the onset of the P-wave due to a high-noise level, we sample P-trigger times from a half-normal distribution with a standard deviation of 2 seconds centered at the predicted P arrival time and S triggers from a normal distribution with a standard deviation of 2 seconds centered on the predicted S arrival time. The shaking amplitude of the triggered phone is set to the value sampled from the Cua & Heaton (2009) relationship.

Figure 3 shows a comparison of trigger times recorded by MyShake devices during the 2016 M5.2 Borrego Springs event in southern California and those simulated by the trigger generation algorithm. Here we assume 0.1% of the population are MyShake users. Figure 3 shows the triggers generated by the P- and S-waves, in addition to some random background triggers in both cases. It is clear that the simulation is able to capture the general characteristics of how the MyShake network responds to events. However, our simple amplitude-based approach loses more P-wave triggers at further distances than the ANN algorithm used by MyShake phones.

During the development of the triggering mechanism, we also attempted use of the Southern California Earthquake Center (SCEC) Broadband platform (Dreger *et al.*, 2015; Maechling *et al.*, 2015) to generate earthquake waveforms from M4.0 to M8.0 at a very dense grid-station configuration, and evaluated the ANN algorithm's triggering performance. However we observed that the SCEC Broadband platform doesn't generate realistic high-frequency P wave components to trigger the algorithm. We therefore proceed with our simulations using the amplitude-based triggering.

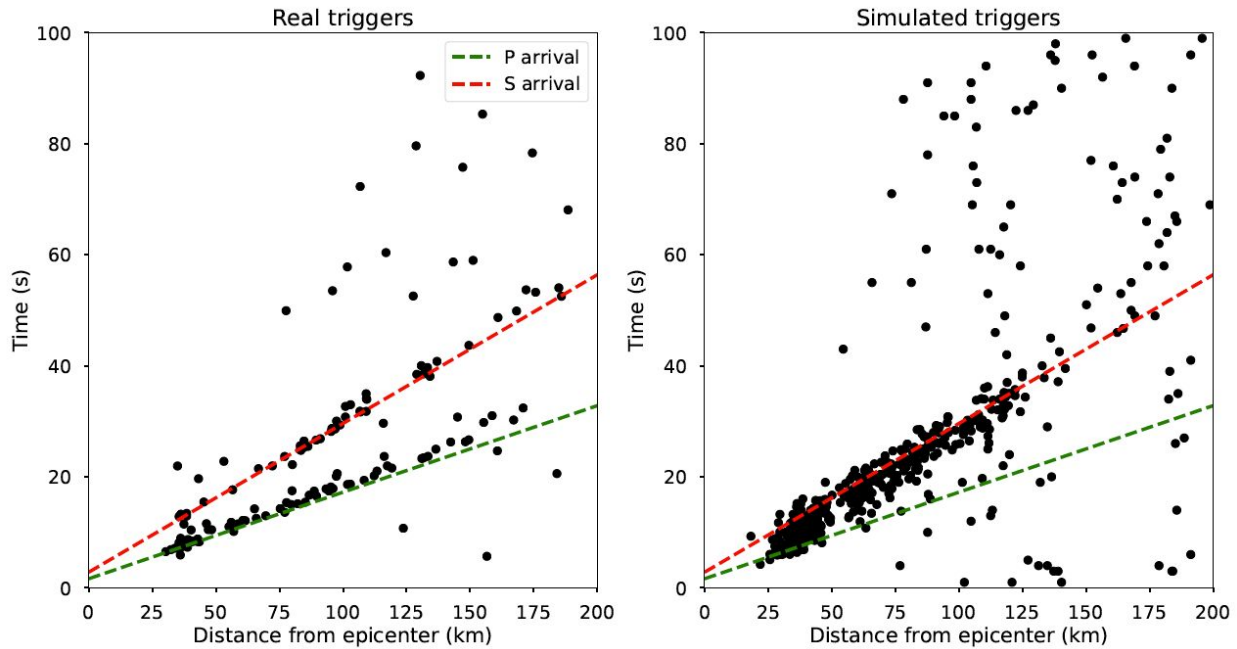

**Figure 3**. Actual and simulated trigger times as a function of epicentral distance for the June 2016 M5.2 Borrego Springs earthquake. The red and green lines show the predicted arrival times of the P and S phases assuming constant velocities of 6.10 and 3.55km/s and an event depth of 10km. These three parameters are fixed in all simulations.

*Network detection with modified DBSCAN clustering*

Our network detection workflow has the task of using either simulated or real-world trigger information to quickly detect that an earthquake is occurring and then determine its origin time, magnitude and hypocenter parameters. MyShake networks present unique challenges for rapid network detection when compared to those composed of traditional seismometers (Kong, Lv, *et al.*, 2019). These include the fact that the network configuration varies over time, the fact that triggers can occur on either P- or S-waves, and potential inaccuracies in trigger timing data due to an inaccurate phone clock and the high noise floor on mobile accelerometers. Finally, the detection algorithm must be capable of accounting for spurious or random triggers that are caused by non-earthquake shaking. The network detection problem essentially is a real-time spatio-temporal clustering problem. We applied a modified version of density-based spatial clustering machine learning algorithm - DBSCAN (Density-Based Spatial Clustering of Applications with Noise) (Ester *et al.*, 1996) to tackle these challenges, which is able to reliably locate earthquakes with reasonable accuracy.

To make the DBSCAN algorithm more reliable and speed up the processing in real-time, instead of using individual phone triggers to search for clusters, we divide the region of interest into grid cells using the Military Grid Reference System (MGRS) (Lampinen, 2001) with 10 by 10 km resolution. Each cell is assigned a weight that can be considered a measure of its reliability in the detection algorithm. The weights are calculated by dividing the number of triggers in each cell by the number of steady phones in the cell, and they are updated over time as more information becomes available. If a cell contains more than five steady phones and the weight is above 0.5, then it is designated a possible candidate for clustering, or we say that the cell is 'activated'.

Once two or more cells are activated within a 20-second sliding window, the DBSCAN algorithm will start to form clusters in order to determine if an event is occurring. The advantages of using DBSCAN are (1) there is no need to specify the number of clusters, (2) the algorithm can automatically label data points that do not belong to any clusters as noise. The DBSCAN algorithm has two parameters: epsilon (a radius parameter) and min_samples (the parameter for setting the minimum number of activated cells to create a cluster). The algorithmic steps are: (1) for each centroid of the activated cells, we draw a 2-dimensional circle of radius epsilon around the centroid. (2) If the number of activated cell centroids inside the circle is larger than the min_samples, we set the center of the circle as the cluster, and all the centroids within the circle belong to this cluster. (3) Loop through all the centroids within the circle with the above two steps to grow the cluster whenever the centroids satisfies the two rules. (4) Centroids that do not belong to any cluster are ignored and treated as noisy outliers. By default we set epsilon at 200 km and min_samples to two grid cells. Once clusters have been formed, each cluster of cells reported by DBSCAN represents a single event. This approach effectively prevents random triggers from being considered part of an earthquake cluster. Furthermore, because DBSCAN is a density-based clustering algorithm that does not require a user-specified number of centroids, the network detection algorithm has the capability of detecting multiple earthquakes simultaneously. This is essential if it is to be run continuously on a global network. A visual explanation of this clustering approach is shown in Figure 4.

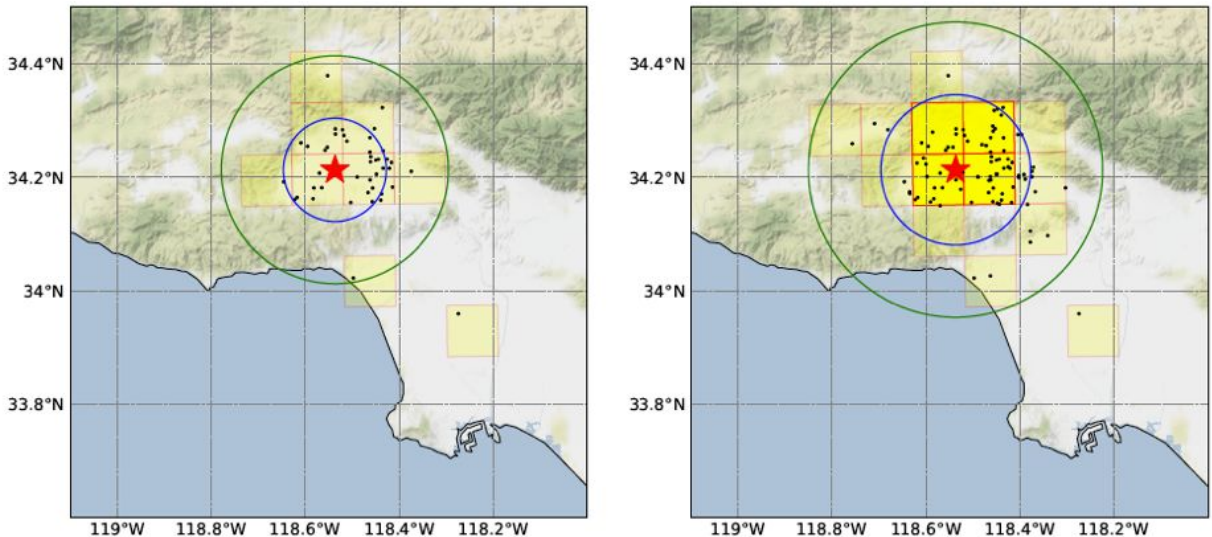

**Figure 4**. Visualization of the detection process during a simulation of the 1994 M6.7 Northridge earthquake in Los Angeles. The panel on the left shows the situation four seconds after the origin time. Black dots represent MyShake devices that have triggered since the start of the event. Yellow squares are the MGRS grid cells known by the algorithm to contain triggers. The green circle represents the estimated location of the P-wave front at this time, while the blue circle shows the location of the S-wave front. An earthquake has not yet been declared because there are insufficient triggers to 'activate' two or more cells. The panel on the right shows the situation at five seconds after the origin time. Four cells (highlighted in yellow) have now 'activated' and they have been clustered to represent a single event. Triggers within these four cells are then used to estimate the event location and magnitude.

The network detection algorithm uses several user-defined parameters that have the potential to exert significant influence on its performance in the real world. These include the minimum number of steady phones required in each MGRS grid cell, the fraction of these steady phones that need to trigger before the cell is considered activated for clustering and the size of the cells themselves. In practice, adjustment of these parameters provides a tradeoff between the speed and accuracy of detection. Typically, the more triggers that must be accumulated before an event is declared, the more accurate the location but the longer it will take to alert. Conversely lowering the threshold for detection will lead to faster alert times but also makes it easier for spurious triggers to influence the detection. Thus, when this network detection algorithm is applied to real MyShake networks, it is likely that the parameters will need to be adjusted from region to region in order to provide optimal results.

*Earthquake location, origin time and magnitude*

Each cluster of cells contains triggers that can be used to locate the event associated with that cluster. This is done by finding a hypocenter location and origin time that minimizes the following objective function, which is a weighted sum of square residual travel times

$$J(X, Y, T) = \sum w_i((t_i - T) - \frac{D_i}{v_{ps}})^2$$

$$D_i = Distance(trigger\ latitude_i,\ trigger\ longitude_i,\ X,\ Y)$$

where w is the weighting of the MGRS cell containing the trigger, t is the trigger time, T is the origin time of the event, D is the distance between the trigger and the event location, and $v_{ps}$ is the velocity of the phase of interest. X and Y above are the event latitude and longitude. The workflow has access to phase information from the triggers. We assume the depth of the event is 10 km without searching for a depth in real time and the goal is to choose a suitable X, Y and T such that this objective function is minimized.

If the minimization fails to converge within 5000 iterations of the Nelder-Mead method (Kelley 1999), a grid search for the optimal location and origin time is carried out. The grid search approach is more time consuming and less accurate due to constraints imposed by the grid step size. However, in practice the optimization fails in less than 5% of all the simulated cases.

Once the event has been located, its distance from each trigger is determined and its magnitude is estimated by providing the distance and ground acceleration value to a random forest regressor trained on synthetic ground accelerations. The training dataset for this model is generated by applying the Cua & Heaton (2009) amplitude relations to a range of synthetic magnitudes and distance values, with magnitudes from M3.5 to M9.0 in steps of 0.1 and distance from 1 to 300 km in steps of 1 km. 1,000,000 samples are generated for P- and S-waves separately. The random forest model encodes the ground motion relationships into a simple map relating epicentral distance and ground acceleration to magnitude. The inputs into the model are the logarithm of epicentral distance and ground motion acceleration, and the output is the estimated magnitude of the earthquake. We performed a grid-search to find the optimal hyperparameters for the random forest model, determining that 100 trees, a minimum number of samples required to split an internal node of 200, and a minimum number of samples required on each leaf of 100 yield the best results.

After testing several approaches, we found that training two separate random forest models for P- and S-wave triggers yields the best results. Separate random forest regressors are trained for magnitude estimation from P- and S-wave amplitude information, with each trigger being passed to the appropriate model according to its associated phase flag. The final event magnitude estimate is then given by the mean of these trigger magnitudes. To test the performance of the trained random forest, we randomly generate accelerations for 100 triggers (a mixture of P and S) for each magnitude from the range of M3.5 to M9.0 with distance randomly sampled from 1 to 100 km. Then we input the acceleration and distance from these 100 triggers at each magnitude to the trained random forest models to estimate the magnitude. The performance of this magnitude estimation approach is elucidated in Figure 5. The models exhibit good performance up to magnitude of about 7.0, with some overestimation at low magnitudes. Figure 5 shows saturation above M7.0, which is expected given the magnitude range of events used to create the Cua & Heaton (2009) relationships.

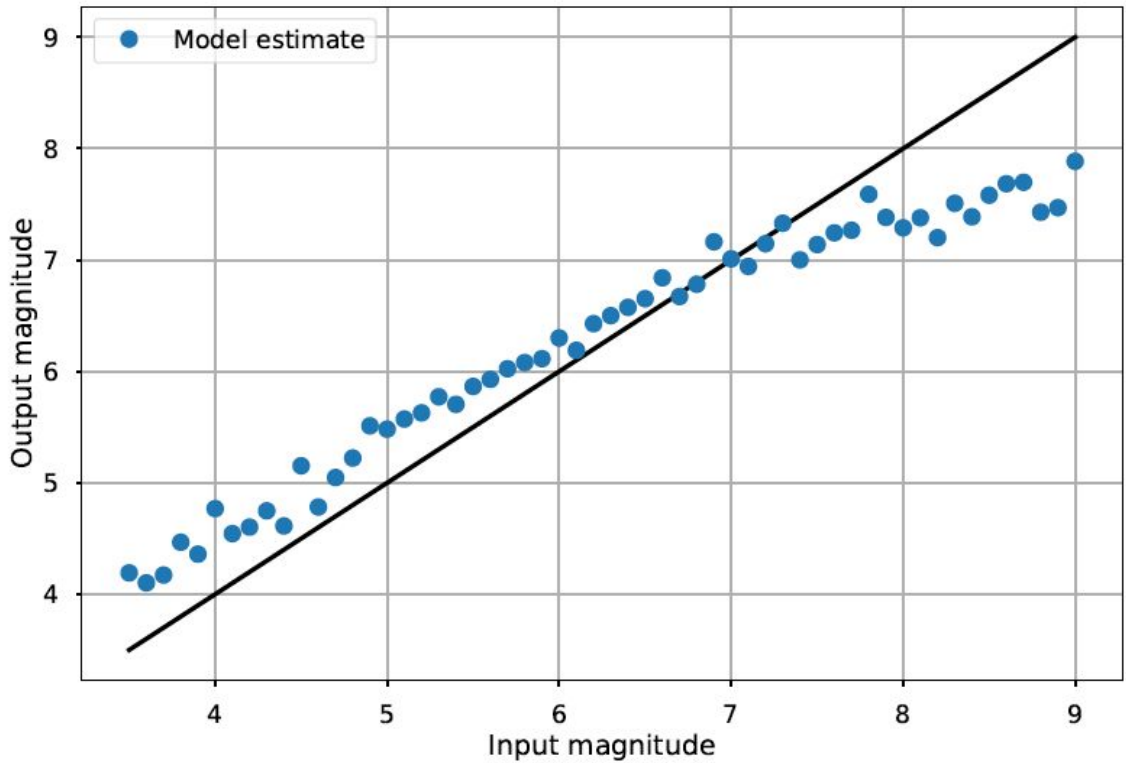

**Figure 5**. Estimated event magnitudes using our magnitude estimation workflow. This test dataset consists of events with magnitudes between 3.5 and 9.0. For each event, 100 trigger distances are drawn from a uniform distribution between 0 and 100 km from the event. Then for each trigger, the Cua & Heaton (2009) distance-amplitude relationships are used to estimate a ground acceleration distribution, from which a value was randomly drawn. The generated P- or S-wave triggers will have a 70% chance of having their phase labeled correctly. Triggers flagged as 'P' are provided to a random forest regressor trained solely on P-wave amplitudes, and triggers flagged as 'S' are provided to a separate regressor trained on S-wave amplitudes. A single magnitude estimate is given for each trigger and the mean of these estimates over all triggers becomes the output magnitude. This workflow exactly emulates the simulation platform.

As time progresses beyond the initial earthquake location step, shaking emanates from the hypocenter in a characteristic pattern governed by the speed of P- and S-waves. In order to improve the initial location estimate, our network detection algorithm can perform a series of updates using additional trigger information as it becomes available. Any additional triggers must be either associated with an earthquake or discarded if they are spurious. The association of new triggers to the detected events is done by checking if the trigger time of the device is within a time-space box for P- and S-waves. This update step is set to occur every 0.5 seconds, with all the triggers associated with the events, although this value could eventually be adjusted dynamically to take population density into account. The location, origin time and the magnitude of the earthquake are updated with the arrival of the new triggers until the user-specified number of updates is reached. The alerting area currently set in the simulation platform is the area with shaking intensity above MMI 4. The shaking intensity is calculated based on the relationship described by Worden *et al.*, (2012).

**Events recorded by the existing MyShake network**

There are currently only a small number of regions where the network of MyShake users is approaching densities sufficient for effective early warning. Only two of these regions, the San Francisco Bay Area and Los Angeles area, have experienced sizable earthquakes since the launch of MyShake in 2016. This in part explains our reasoning for the creation of the simulation platform, which allows us to test hypothetical scenarios around the world. However, as a test of our network detection algorithm, we apply it in simulated real-time to the stream of triggers from actual MyShake-phones returned during the January 4th 2018 M4.4 Berkeley and June 10th 2016 M5.2 Borrego Springs events.

The Berkeley event occurred directly beneath an urban area. We input these triggers into the network detection workflow. Movies S1-3 illustrate the full results, while Figure 6 shows a snapshot at the time of first alert with the initial location and magnitude estimation for this event. Because of the relatively high density of MyShake phones in the city of Berkeley, we found that setting the MGRS grid cell size used for clustering to 1x1 km cells and the minimum number of steady phones required within each cell to two (Figure 6b, Movie S2) increased the warning time compared with their default values of 10 km resolution and six phones (Figure 6a, Movie 1S). With this improvement, the first alert is sent out 5.7 seconds after the origin time of the event, which gives centers of San Francisco and San Jose 0.7 s and 13.3 s warning time respectively (until the predicted S-wave arrival). This illustrates the need to have an adaptive threshold for different regions depending on the density of the network. However, even with the default values the event is detected within 6.7 seconds of the origin time, providing several seconds of warning for much of the San Francisco Bay Area. The initial location has a relatively small epicentral distance error of 4 km, and remains very close to this value during subsequent updates. The magnitude of the event is a little overestimated, as expected given the test shown in Figure 5, and correspondingly the estimated intensities are slightly higher for various locations for both cases. This test suggests that had the network detection algorithm been operational during this event in 2018, it could have provided early warnings.

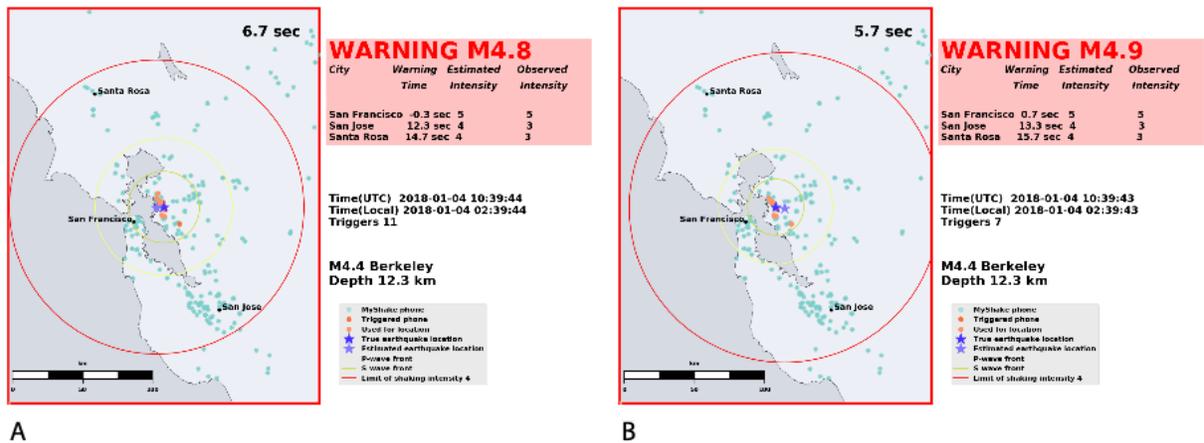

**Figure 6**. Initial performance of the network detection algorithm using real MyShake triggers to detect and locate the January 2018 M4.4 Berkeley event. Both panels correspond to the moment of first location of the event. Panel A shows performance with the default settings of 10 km resolution MGRS grid cells for clustering and a minimum of 6 steady phones required in each cell in order for it to be considered for clustering. Panel B shows the result with parameters modified to optimize detection speed, with 1 km resolution MGRS grid cells and a minimum of two phones steady. Green dots are devices running MyShake at the time of the quake, while orange dots are devices that triggered. The figures also show the estimated positions of the P and S wavefronts at the shown snapshot in time, and the estimated radius of shaking intensity greater than MMI 4 (red circle). When optimized for detection speed, the algorithm locates the event using 7 triggers within 5.7 seconds of the origin time, providing warning for much of the San Francisco Bay Area. Movies S1 and S2 should be consulted for more information.

The Borrego Springs event poses a more challenging test of the network detection algorithm because it occurred in a remote location about 50 km south of Palm Springs, where all of the initial triggers are located. Thus the initial azimuthal distribution of triggers is not ideal and the algorithm must be capable of associating later triggers to the same event, even though they occur at great distances from the original cluster. It is important for any network detection workflow to deal with such a situation, since it will be common in regions featuring major faults far from population centers. Despite these challenges, using its default settings, our network detection algorithm performs relatively well, locating the event with an initial error of about 14 km and a magnitude underestimation of 0.3 units. 5.5 seconds of warning time is provided to Palm Springs and people near San Diego would receive warning of about 40 seconds. The red circle shows the radius of the region expected to experience shaking of intensity 4 and above.

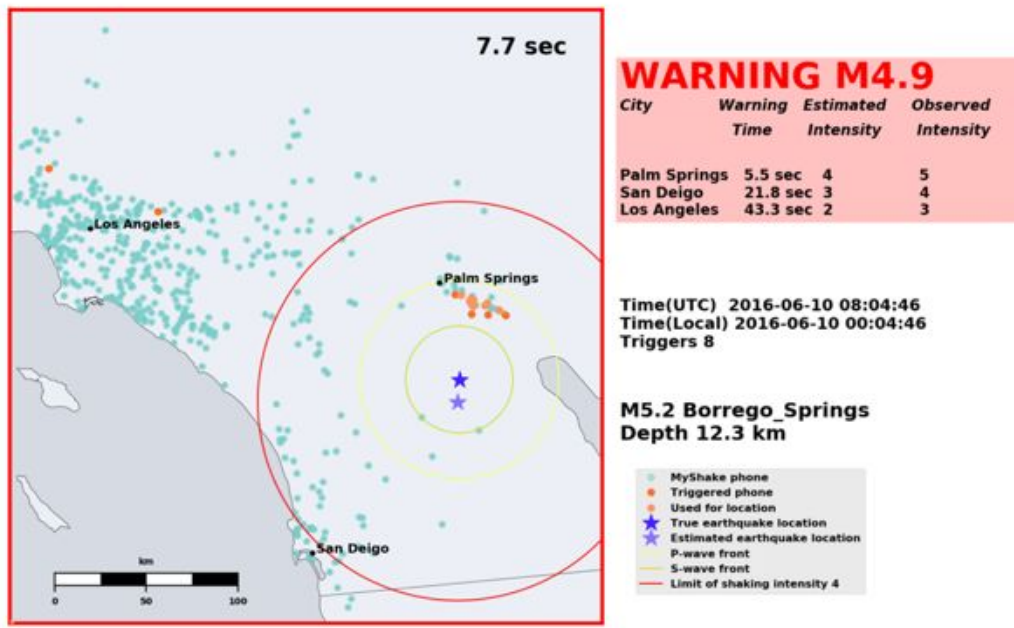

**Figure 7**. Performance of the network detection algorithm using real MyShake triggers from the M5.2 Borrego Springs earthquake. This panel shows the network detection algorithm's performance at the moment of first alert. 8 triggers, all of which appear to occur on arrival of the P-wave in the Palm Springs area, are used to initially locate the event.

**Discussion**

Our network detection algorithm and the simulation platform are designed to facilitate global earthquake early warning capabilities for the MyShake smartphone seismic network. Once the current algorithm is deployed and more real MyShake trigger data becomes available, we foresee both challenges and opportunities for improvement.

The current triggering mechanism for individual phones in our simulation platform is an amplitude-based approach, which captures the general triggering pattern from the current MyShake network but could be improved. The artificial neural network (ANN) algorithm used in the MyShake application uses both the frequency and amplitude information from the waveforms. This is different from the current implementation of our amplitude-based approach and represents a future option for improving the simulation platform.

Rapid and accurate magnitude estimation is another area where improvements could be made. The challenge is that earthquakes of a given magnitude produce broad distributions of ground motion with large uncertainties at a given distance. This is especially true for smartphones, which are typically in buildings and exhibit a wide range of ground-coupling scenarios. There is also the added complication that the phones can be triggered either on the P- or the S-phase. In the simulation platform, our trigger generation workflow attempts to account for some of the uncertainties by sampling from a distribution with uncertainties built-in. However, since the peak ground motions for earthquakes of different magnitudes overlap considerably at a given distance, especially for the P-wave amplitude, it is difficult for any model to accurately estimate magnitude from an initial acceleration observation alone.

The two real events that recorded by MyShake users show very interesting results and provide insight to potential improvements to the system. The Berkeley event re-run illustrates the need to have an adaptive detection procedure where the parameters of the algorithm can be updated to reflect different network configurations. As indicated by Figure 6, the effectiveness of the warning can change with different parameter settings: regions with higher population density can make use of smaller MGRS grid cells to improve the detection speed of the earthquake, thus increase the warning time and reducing the radius of the blind zone for the region. The Borrego Springs event (Figure 7) illustrates a case when there are few users close to the earthquake, meaning that it takes a relatively long time for the system to detect the earthquake. In California, there is also a well established, relatively dense network of traditional seismic sensors that could potentially be utilized alongside MyShake users to facilitate faster and more reliable EEW in this region.

**Conclusion**

The global MyShake network is currently in its infancy and there are an insufficient number of users in most parts of the world to reliably evaluate the earthquake early warning capabilities of the system. For this reason, we have built a simulation platform and used it to develop a new network detection algorithm.

Two of the earthquakes recorded by MyShake users are used to evaluate the performance of the network detection algorithm on real phone data: The January 4th 2018 M4.4 Berkeley event and the June 10th 2016 M5.2 Borrego Springs event. Despite the challenges associated with the currently sparse MyShake network, both events would have been located rapidly and relatively accurately using our network detection approach. Several seconds of warning could thus have been provided to major urban areas before the onset of the largest shaking.

Part two of this series builds on the work presented here to use the simulation platform and the network detection algorithm to evaluate the first order performance of MyShake in hypothetical earthquake scenarios around the world. These simulations provide us with an understanding of the potential future performance of MyShake networks.

**Data and Resources**

The USGS Comcat catalog can be accessed at: https://earthquake.usgs.gov/fdsnws/event/1/. The data for the Gridded Population of the World can be accessed at https://beta.sedac.ciesin.columbia.edu/data/set/gpw-v4-population-count-adjusted-to-2015-unwpp-country-totals. MyShake data are currently archived at Berkeley Seismology Lab and use is constrained by the privacy policy ( http://myshake.berkeley.edu/privacy-policy/index.html ).

**Acknowledgements**

The Gordon and Betty Moore Foundation fund this analysis through grant GBMF5230 to UC Berkeley. The California Governor's Office of Emergency Services (Cal OES) fund this analysis through grant 6142-2018 to Berkeley Seismology Lab. We thank the previous and current MyShake team members: Roman Baumgaertner, Garner Lee, Arno Puder, Louis Schreier, Stephen Allen, Stephen Thompson, Jennifer Strauss, Kaylin Rochford, Akie Mejia, Doug Neuhauser, Stephane Zuzlewski, Asaf Inbal, Sarina Patel and Jennifer Taggart for keeping this Project running and growing. All the analysis of this project is done in Python, particularly the

ObsPy package (Beyreuther et al., 2010; Wassermann et al., 2013; Krischer et al., 2015). We also thank all the MyShake users who contribute to the project!